Thermal Processing of Silicate Dust in the Solar Nebula: Clues from Primitive Chondrite Matrices

Edward R. D. Scott and Alexander N. Krot

Hawai'i Institute of Geophysics and Planetology, University of Hawai'i at Manoa,

Honolulu, Hawai'i 96822, USA

Corresponding author: Edward R. D. Scott

escott@hawaii.edu

ph:808-956-3955; fax:808-956-6322

Short title: Nebular Dust in Primitive Chondrite Matrices

Abstract:

The most abundant matrix minerals in chondritic meteorites—hydrated phyllosilicates and ferrous olivine crystals—formed predominantly in asteroids during fluid-assisted metamorphism. We infer that they formed from minerals present in three less altered carbonaceous chondrites that have silicate matrices composed largely of micrometer- and nanometer-sized grains of crystalline forsterite, $Mg_2SiO_4$, and enstatite $MgSiO_3$, and amorphous, ferromagnesian silicate. Compositional and structural features of enstatite and forsterite suggest that they formed as condensates that cooled below 1300 K at ~1000 K/h. Most amorphous silicates are likely to be solar nebula condensates also, as matrix, which is approximately solar in composition, is unlikely to be a mixture of genetically unrelated materials with different compositions. Since chondrules cooled at 10-1000 K/h, and matrix and chondrules are chemically complementary, most matrix silicates probably formed close to chondrules in transient heating events. Shock heating is favored as nebular shocks capable of melting millimeter-sized aggregates vaporize dust. The crystalline and amorphous silicates in the primitive chondrite matrices share many characteristic features with silicates in chondritic interplanetary dust particles suggesting that most of the crystalline silicates and possibly some amorphous silicates in the interplanetary dust particles are also nebular condensates. Except for small amounts of refractory oxides from the inner CAI-forming region and presolar dust, most of the crystalline silicate dust that accreted into chondritic asteroids and long-period comets appears to have formed from shock heating at ~2-10 AU. Forsterite crystals around young stars may have a similar origin.

Subject headings: astrochemistry—minor planets, asteroids—solar system: formation—planetary systems: protoplanetary disks—comets: general —meteors, meteoroids, meteorites

# 1. INTRODUCTION

The components in chondrites – chondrules, refractory inclusions, and matrix material -- provide an extraordinarily detailed record of the processes that transformed the composition and mineralogy of dust in the protoplanetary disk before planetesimals accreted.  Although much attention has focused on the chondrules and refractory inclusions, it is the matrix material that contains the most information about the origin and evolution of the dust in the disk.  Matrix is a mixture of 10 nm to 5 μm grains of silicates, oxides, sulfides and other minerals that coats chondrules and refractory inclusions and partly fills the interstices between them (see Scott & Krot, 2003).  Matrix material has a bulk chemical composition close to that of the whole chondrite with near-solar elemental proportions (Brearley 1996) and contains traces of two types of presolar material, interstellar organic material and circumstellar inorganic grains (Alexander 2001; Zinner 2003).  Matrix grains coated chondrules and refractory inclusions as these components were accreting in the protoplanetary disk (Cuzzi 2004). Because of its fine grain size, high porosity and permeability, matrix is much more susceptible to alteration by aqueous fluids and heating in asteroids than the other chondritic components. As a result, it is difficult to analyze the complex mineral intergrowths in matrices and to distinguish altered phases from pristine ones.

The origins of chondrules and refractory inclusions in transient heating events in the protoplanetary disk are relatively well understood, though many details including the heat source remain obscure (e.g., Zanda 2004; Wood 2004).  Refractory inclusions formed in the presence of $^{16}$O-rich gas, probably in the innermost portion of the Sun's rapidly accreting disk, by evaporation and condensation at high ambient temperatures. Chondrules formed later in cooler, more distant regions in $^{16}$O-poor gas by melting of solids with relatively minor vaporization and condensation.  The origins of matrix material, however, have remained obscure mainly because of difficulties in distinguishing nebular and asteroidal characteristics (Scott et al. 1988; Brearley 1996). Anders and Kerridge (1988) noted that, "controversy permeates every aspect of the study of matrix material, from definition of matrix to theories of its origin."  According to Brearley (1993), "there are almost as many proposed theories for the origin of matrix materials as there are studies."

The role of impacts in forming matrix has been especially controversial. Transmission electron microscopy by Brearley (1993) has shown that matrix rims did not form in hypervelocity impacts on asteroids, as Bunch et al. (1991) suggested, and studies of irradiation tracks by Metzler (2004) and in situ analyses of noble gases by Nakamura et al. (1999) have demonstrated that the fine-grained matrix rims around chondrules were not acquired in asteroidal regoliths, as Symes et al. (1998) concluded.  Nevertheless, matrix material has been modified by impacts as many chondrites contain interstitial mixtures of matrix material and fragments of chondrules and other components, and some chondrites are breccias of rock fragments that were processed at different depths in their parent asteroid.  Thus, asteroidal impacts, like alteration and heating, have conspired to obscure the primordial features of matrix material.

Matrix material has traditionally been considered as the fine-grained portion of chondrites that equilibrated with the nebula at low temperatures (Larimer and Anders 1967). However, this view originated when hydrated silicates, magnetite, carbonates and other minerals that are abundant in the matrices of many carbonaceous chondrites were thought to have formed in the disk.  Since these phases are now known to have formed during aqueous activity on asteroids (Brearley 2003), the concept of matrix as a low-temperature component needs to be reassessed. The presence of diverse materials including organics, refractory silicate grains, and presolar minerals suggests that it is a complex mixture of materials that formed under diverse conditions and accreted at low temperatures with ice and other volatiles (Scott & Krot 2003).

The minerals observed in the matrices of different chondrites are extraordinarily diverse, as they appear to reflect a wide variety of nebular and asteroidal processes (Table 1). To disentangle the effects of these processes on matrix minerals, a holistic approach is needed to integrate what has been learned about asteroidal alteration from studies of the various chondritic components (Scott & Krot 2003). Our goal here is to identify matrix minerals that may be products of solar nebular processes and to understand how they formed, and how they are related to chondrules and silicates in chondritic interplanetary particles (IDPs), which probably originate in comets (Bradley 2003). We infer that the fine-grained silicate material in chondrite matrices is remarkably similar to the fine-grained silicate material in IDPs. Constraints derived from mineralogical studies of matrix and IDPs suggest that much of the dust that accreted in the protoplanetary disk experienced brief, localized, nebular heating events like those that produced chondrules.

2. CHONDRITE MATRIX MINERALS
2.1. *Overview*

In this section, we review what is known about the thermal and alteration history of chondrites with diverse minerals in their matrices (Table 1) to identify pristine chondrites with the least altered and least metamorphosed matrices. The diverse matrix minerals formed during nebular processes that generated the different chondrite groups (CO, CM, CV etc) and asteroidal processes that produced the different petrologic types (1-6). The nature and extent of asteroidal modification of the components in chondrites depended on the temperature and duration of metamorphism and the supply of aqueous fluid. As can be inferred from their name, type 1 chondrites, which are composed almost entirely of matrix, were originally thought to be the most pristine chondrites as they are closest in composition to the solar photosphere (e.g., Palme 2003). But they are now known to be the most heavily affected by aqueous alteration (e.g., Scott & Krot 2003). Type 2s are less altered than type 1s, and types 4-6 were strongly heated to 900-1300 K. Type 3 chondrites are the least altered and least metamorphosed chondrites.

The least altered or heated type 3 chondrites, which are called type 3.0, are found in the CB group, but these unfortunately lack matrix material (except for a few altered rock fragments), and appear to have formed late when the chondrules condensed and accreted in a dust-free nebula (Krot et al. 2002; Amelin et al. 2004). The least altered type 3 chondrites with matrices are a few type 3.0 chondrites in the CO and LL groups, and some ungrouped type 3 carbonaceous chondrites. The CV3 chondrites, which include the most studied chondrite, Allende, cannot be arranged in a simple metamorphic sequence like the CO3 and LL3 chondrites, but the degree of disorder in the graphitizable carbon suggests they should be classed as type 3.1 to >3.6 (Quirico et al. 2003; Bonal et al. 2004).

The two most abundant matrix minerals are phyllosilicates, which contain water, and iron-rich olivine. Although their origin has been controversial, there is now much evidence against a nebular origin for both minerals.

2.2. *Phyllosilicates*

Matrices and other components in types 2 and 3 chondrites commonly contain hydrated phyllosilicates such as saponite, serpentine, and smectite. They typically occur with hydrated oxides like ferrihydrite (FeOOH·$n$H$_2$O) and sulfides like tochilinite [6Fe$_{0.9}$S·5(Fe,Mg)(OH)$_2$)]. Although nebular formation mechanisms for phyllosilicates have been proposed (Ciesla et al. 2003), petrologic studies of chondritic components including those that are readily identified in many groups (like type I chondrules) strongly suggest that phyllosilicates formed predominantly in asteroids (Brearley 2003).

2.3. *Ferrous Olivine*

The most abundant phase in the matrices of many type 3 chondrites is ferrous olivine, (Fe$_x$Mg$_{1-x}$)$_2$SiO$_4$, which has an Fe/(Fe+Mg) ratio ($x$) of ~0.3-0.6 and a grain size of <100 nm to 20 μm. A variety of nebular and asteroidal origins have been proposed for ferrous olivine in chondrites (Brearley 2003): reactions at high temperatures (>1200 K) in a highly oxidized nebula (Dohman et al. 1998; Weisberg and Prinz 1998), annealing of amorphous silicate in the nebula (Brearley 1993), dehydration of phyllosilicates in asteroids (Krot et al. 1997; Brearley 1999), growth from aqueous fluids during heating (Krot et al. 2004a). The poor crystallinity, abundant voids, and inclusions of poorly graphitized carbon and pentlandite [(Fe,Ni)$_9$S$_8$] in ferrous olivine argue against a high temperature origin (Brearley 1999; Abreu and Brearley 2003). Formation of ferrous olivine in the solar nebula from enstatite (MgSiO$_3$) and metallic Fe,Ni at low temperatures is kinetically inhibited (Palme & Fegley 1990).

The distribution of ferrous olivine in matrices, chondrules and refractory inclusions in CV3 and CO3 chondrites strongly suggests that it formed in the parent asteroids during heating in the presence of an aqueous fluid (Krot et al. 2004a). In CV3 chondrites, the size and abundance of ferrous olivine crystals increases with increasing degree of order in the graphitizable carbon. Thus Allende, which is a type >3.6, contains the most ordered carbon (Bonal et al. 2004) and the largest Fe-rich, matrix olivines -- up to 20 μm (Brearley 1999). At the opposite extreme, Kaba, which is a type 3.1, contains the smallest olivines ~1 μm (Keller & Buseck, 1990) and the least ordered carbon. CV3 chondrites also contain crosscutting ferrous olivine veins that clearly formed after accretion. In CO3 chondrites, the amount of ferrous olivine in amoeboid olivine aggregates, which are refractory inclusions made of CAIs and olivine, is correlated with the degree of metamorphism experienced by the chondrite. Olivine in aggregates in type 3.0 chondrites contains nearly pure Mg$_2$SiO$_4$, forsterite, with Fe/(Fe+Mg) ~0.01, whereas in type 3.8, the Fe/(Fe+Mg) ratio is ~0.35 (Chizmadia et al. 2002). In CV3 chondrites, the Fe-rich olivine is associated with other minerals that are characteristic of altered chondrites such as Ca-Fe-rich pyroxene (CaFe$_x$Mg$_{1-x}$Si$_2$O$_6$) and magnetite Fe$_3$O$_4$. Oxygen isotopic analyses of these phases show large mass-dependent fractions that are characteristic of low-temperature alteration processes (<700 K) (see Krot et al. 2004a). Thus there is good evidence that matrix ferrous olivine formed predominantly in asteroids.

2.4. *Matrices of Pristine Chondrites*

If we exclude the altered chondrites in Table 1 that have matrices and other components containing abundant phyllosilicates and/or ferrous olivine, there are four type 3 chondrites that appear to have the least altered components. Three are carbonaceous chondrites -- ALHA77307, which belongs to the CO group, and Acfer 094 and Adelaide, which are ungrouped chondrites related to the CM and CO groups. The fourth, Kakangari, is a member of the K group, which has links to enstatite and other chondrites. ALHA77307 is the least metamorphosed, type 3.0 chondrite in the CO group and Acfer 094 has a comparable thermal history (Grossman 2004). Acfer 094 and Adelaide both lack phyllosilicates in their matrices (Greshake 1997; Brearley 1991) and have among the best preserved CAIs of any chondrite (Huss et al. 2003; Krot et al. 2004b). Matrices in Acfer 094 and Kakangari, though not Adelaide, lack ferrous silicates (Brearley 1989, 1991).

Matrices in the four pristine chondrites consist largely of micrometer-sized or smaller crystalline silicates, which, except in Kakangari, are embedded in amorphous material. The amorphous material is largely composed of Si, Mg, Fe, and O and in ALHA77307 it forms 1-5 µm units, which are very heterogeneous on the micrometer-scale (Brearley 1993; Greshake, 1997). The amorphous materials in Acfer 094 and ALHA77307 have comparable compositions. Relative to the bulk matrix, the amorphous material is depleted in Ti, Fe, and Mg and enriched in Si, Ni, Cr, Mn, P and S by factors of ~0.5-2. Concentrations of Al and Ca range around the bulk matrix composition.

The crystalline silicates in all four chondrite matrices are mostly 100-1000 nm crystals of magnesian olivine and low-Ca pyroxene ($Fe_xMg_{1-x}SiO_3$) with Fe/(Mg+Fe) ratios of <0.05 (Table 2). They typically occur as isolated, rounded or elongated grains, but clusters and aggregates also occur (Fig. 1) and are especially common in Kakangari. Olivine is generally more abundant than low-Ca pyroxene (except in Kakangari), and both may have high concentrations of Mn (0.6-2 wt.% MnO). [Olivines in chondrules with similarly low concentrations of Fe typically contain <0.1 wt.% MnO (Brearley & Jones 1998)]. Many low-Ca pyroxene crystals show intergrowths of ortho- and clino-pyroxene indicating that they were cooled at ~1000 K/h from the protopyroxene stability field above 1300 K (Brearley 1993; Greshake 1997). Ferrous olivine crystals <300 nm in size are also present in ALHA77307. Unlike the magnesian silicates, which are highly crystalline, the ferrous olivines are poorly crystalline and they probably formed from the amorphous phase at low temperatures, like the nanoscale phyllosilicates (Brearley 1993). Table 2 summarizes the properties of the magnesian silicates and the amorphous phase in the pristine C chondrites. In addition, there are about 5 vol. % of rounded grains of Ni-bearing sulfides, mostly pyrrhotite, $Fe_{1-x}S$, which are generally 100-300 nm in size and heterogeneously distributed in the amorphous material. In ALHA77307, Fe,Ni grains 25-200 nm in size are also present (Brearley, 1993).

The matrix in the K chondrite, Kakangari, differs from the C chondrite matrices in that feldspar crystals — albite ($NaAlSi_3O_8$) and anorthite ($CaAl_2Si_2O_8$) — are present, amorphous silicate is absent (Brearley 1989). The Kakangari matrix is also unusual in that its mineralogy and chemical composition are surprisingly similar to those of the Kakangari chondrules, and ferrous iron in matrix and chondrules has been largely reduced to metallic Fe. Olivine is generally enclosed within pyroxene, which is more abundant and has a slightly higher Fe/(Fe+Mg). Besides the magnesian silicates, the matrix contains 10 vol.% albite crystals 250-1000 nm in size, which are commonly intergrown with olivine and pyroxene in 2-8 µm particles, and 1 vol.% anorthite grains up to 500 nm in size, which are intergrown with olivine in granular particles (Brearley, 1989). Albite was probably cooled rapidly to preserve the high-temperature form, and the pyroxene shows ortho-clino intergrowths indicative of cooling at ~1000 K/h below 1300 K, as in the three other pristine chondrites. The matrix mineralogy implies formation over a range of nebular temperatures: forsterite and forsterite-anorthite particles forming at higher temperatures than the albite-forsterite-enstatite and forsterite-enstatite particles (see Yoneda & Grossman, 1995). Kakangari has been metamorphosed mildly but the matrix minerals do not appear to have been modified significantly.

3. DISCUSSION

Out of ~4000 individual chondrites, only four have matrices that are relatively unaffected by mineralogical changes due to aqueous alteration and/or metamorphism (Table 2). They all contain significant proportions of enstatite and forsterite (~10-70%), which are 100-1000 nm in size. Matrices of the three carbonaceous chondrites are similar and all contain amorphous Fe-Mg-Si-O material. The matrix of the K chondrite, Kakangari, lacks amorphous material and formed from a distinctly different dust mixture . Although other chondrites have matrix minerals that are very different (Table 1), it is plausible that they formed from precursor material that was composed largely of the minerals in the matrices of the four pristine chondrites. Amorphous material is found in CM chondrites (Barber 1981; Chizmadia & Brearley 2003) and ordinary chondrites (see Scott & Krot 2003). Manganese-rich, magnesian olivines and pyroxenes are also present in the matrices of LL3 and CM2 chondrites (Klöck et al. 1989). Amorphous silicate could have been readily converted to ferrous olivine or phyllosilicates when heated in asteroids in the presence of water.

Silicates that are relatively ferroan (Fe/(Fe+Mg) > 0.1) are absent in the matrices of the most pristine carbonaceous chondrite, Acfer 094, in Kakangari, and probably the enstatite chondrites (EH and EL). Ferroan olivine crystals in the matrices of ordinary chondrites (H, L, and LL) appear to be mostly alteration products, but the possibility that some formed in the nebula cannot be excluded. Since fine-grained matrices were most susceptible to alteration in asteroids, it is hardly surprising that chondrites with pristine matrices are ten times rarer than Martian meteorites.

3.1 *Comparison of chondrite matrices, chondritic interplanetary dust particles, and cometary dust*

The matrices of the least altered, carbonaceous chondrites are remarkably similar in mineralogy to the carbon-free portions of chondritic porous interplanetary dust particles (IDPs) and the silicate grains inferred to be present in outflowing coma dust from long-period comets like Hale Bopp (Hanner 1999; Bradley 2003; Brownlee 2003; Wooden et al. 2004). The silicate portions of primitive carbonaceous chondrite matrices, chondritic IDPs, and comets are all largely composed of forsterite and enstatite with Fe/(Fe+Mg) <0.05 and amorphous, heterogeneous Fe-Mg-Si-O material (Table 2). Chondrite matrices and chondritic IDPs both contain sub-micrometer, Mn-rich forsterite and enstatite crystals (Klöck et al. 1989), though IDPs appear to have higher proportions of enstatite (Bradley 2003). IDPs, like primitive matrices, contain enstatite crystals with ortho-clino intergrowths characteristic of rapid cooling at ~1000 K/h from 1300 K (Bradley et al. 1983). Amorphous particles in chondrite matrices are chemically heterogeneous and may contain rounded grains of metal and sulfide (Brearley 1993). We infer that some may be related to particles in IDPs made of glass with embedded metal and sulfides called GEMS. However, GEMS are 10 × smaller and typically have more magnesian glass (Bradley 2003; Keller and Messenger 2004).

Many differences between the matrices of primitive carbonaceous chondrites and chondritic IDPs can readily be attributed to formation at different heliocentric distances. Concentrations of C (4-45%), and presolar grains (~900 ppm) are ~10-100 × higher in chondritic IDPs than in chondrite matrices (Keller et al. 1994a; Messenger et al. 2003; Nguyen & Zinner 2004a, b; Nagashima et al. 2004). Other differences, such as the higher proportion of enstatite to forsterite in IDPs (Bradley 2003) relative to carbonaceous chondrite matrices (Table 2), probably reflect differences in thermal processing. Nevertheless, the overall resemblance suggests that similar processes helped to form the crystalline silicate particles in the matrices of carbonaceous chondrites and the chondritic IDPs, as well as some fraction of the amorphous particles. The origins of these materials are discussed in sections 3.2 (amorphous silicates) and 3.3 (crystalline silicates); cometary silicates are discussed further in section 3.5.

The resemblance between the matrices of pristine, carbonaceous chondrites and cometary dust contradicts the general belief that the crystalline and amorphous silicates in comets and chondritic IDPs have no counterparts in meteorites (e.g., Hanner 1999; Bradley et al. 1999). Because of this close resemblance, it might be argued that pristine carbonaceous chondrites come from comets. However, this appears very unlikely as ALHA77307, for example, comes from a body that was heated internally to ~800 K to make the type 3.8 CO chondrites. The similarity between silicates in chondrite matrices and comets is probably yet another indication that these bodies did not form in fundamentally different ways, and that the ranges of many of their properties may overlap. The existence of pristine, matrix-rich carbonaceous chondrites that were not aqueously altered, like Acfer 094, suggests that some parts of asteroids were never heated significantly, consistent with the discovery of comet-like activity from a C class asteroid belonging to the Themis family (Hsieh et al. 2004).

### 3.2. *Origins of Amorphous Silicate*

Amorphous particles in matrices and IDPs are derived from multiple sources. Two of 40 GEMS in chondritic IDPs that were analyzed by Messenger et al. (2003) were isotopically anomalous and formed around other stars, but the remainder had normal oxygen isotopic compositions, i.e., within experimental error of ±40‰ from the terrestrial value. The latter could have formed in the solar nebula or they may represent grains that were isotopically homogenized in the interstellar medium. Presolar GEMS-like particles have also been reported in chondrite matrices by Nguyen & Zinner (2004a) and Mostefaoui et al. (2004). Although there are few detailed O-isotopic studies of matrix, the bulk chondrite and mean chondrule compositions are similar showing that matrices are not $^{16}$O-rich like refractory inclusions. (Kakangari's matrix has been analyzed and was found to be slightly lighter in its O-isotopic composition than its chondrules (Prinz et al. 1989).) Thus most matrix particles probably formed in $^{16}$O-poor regions, like chondrules. Because CAIs formed at high ambient temperatures near to the protosun in an $^{16}$O-rich region and unlike chondrules do not contain glass (Wood 2004), amorphous materials are unlikely to have formed in the CAI-forming region. Amorphous particles (GEMS) that are $^{16}$O-rich, like CAIs, have been reported in IDPs (Engrand et al. 1999) and may also be present in chondrite matrices. They may be grains that were isotopically homogenized prior to infall into the disk, as the mean oxygen isotopic composition of the pre-solar silicates probably matches that of the CAIs (Scott & Krot 2001; Clayton 2002; Yurimoto & Kurimoto 2004). We suggest that GEMS that formed in the solar nebula might be distinguished from presolar, but isotopically normal GEMS with more accurate oxygen isotopic analyses. Keller & Messenger (2004) analyzed the chemical composition of 144 GEMS and infer that ~10-20% have near solar chemical compositions and may have been isotopically and chemically homogenized in the interstellar medium.

Amorphous silicate particles may form by irradiation or destruction of crystals by shock processes (e.g., Bradley 2003; Brucato et al. 2004), or by condensation if crystal growth is kinetically inhibited (Gail 2003). Radiation-damaged grains are found in IDPs (Bradley et al. 1999) but are very rare in chondrites except for asteroidal regolith samples (Zolensky et al. 2003). The amorphous and crystalline components in matrix, which differ in chemical composition, are likely to be genetically related as bulk matrix is approximately solar in composition. Since most crystalline silicates in matrix are probably nebular condensates (see below), the amorphous matrix silicates probably also condensed in the nebula.. They may have formed under conditions where condensation of crystalline magnesian silicates and metal grains was prevented by rapid cooling rates and/or low pressures.

3.3. *Origins of Crystalline Magnesian Silicates*

Given the low concentrations of presolar crystalline silicates in chondritic matrices and IDPs, their abundant crystalline magnesian silicates must be solar nebula products. Two possible origins have been discussed for magnesian silicates in matrices and IDPs: annealing of amorphous material (e.g., Brearley 1993; Nuth et al. 2000) and nebular condensation (Bradley et al. 1983; Klöck et al. 1989; Greshake 1997). Annealing experiments show that forsterite (and tridymite) may crystallize from amorphous Mg-Si-O smoke (Hallenbeck et al. 2000; Nuth et al. 2000; Fabian et al. 2000; Thompson et al. 2003). Such an origin has been proposed for the Kakangari matrix, where nearly all Fe is reduced to Fe,Ni (Brearley 1989). However, the wide mineralogical variations among the Kakangari matrix particles are more consistent with condensation over a wide range of temperatures. For chondritic matrices and IDPs, crystallization of nearly pure $MgSiO_4$ and $MgSiO_3$ from amorphous Fe-Mg-Si-O material appears unlikely as olivine and pyroxene readily accommodate $Fe^{2+}$ (Brownlee 2003).

Direct evidence in favor of a condensate origin for the magnesian silicates in chondrite matrices and IDPs comes from their composition and structure. Nearly pure crystals of $Mg_2SiO_4$ and $MgSiO_3$ can form in the solar nebula at ~1300-1400 K, assuming a total pressure of $10^{-3}$ atm (Yoneda and Grossman 1995). The common occurrence of separate enstatite and forsterite crystals suggests that forsterites were partially isolated during condensation so that enstatite did not form from forsterite (Petaev & Wood 1998). High Mn concentrations found in some magnesian olivines and pyroxenes in both matrices and IDPs are consistent with condensation as Mn should condense into these minerals at ~1200 K (Wai & Wasson 1977; Klöck et al. 1989). However, detailed studies of Mn condensation under equilibrium and disequilibrium conditions are needed. Supporting evidence for a condensation origin for the magnesian silicates in IDPs and matrices comes from analyses of amoeboid olivine aggregates in pristine chondrites (Krot et al. 2004c; Weisberg et al. 2004). These irregularly-shaped aggregates of forsterite and CAI minerals have grain sizes of 1-20 μm and formed by condensation in the solar nebula. Their olivines and pyroxenes are chemically similar to matrix magnesian silicates: e.g., Fe/(Fe+Mg) values of 0.002-0.02 and up to 1.3 wt.% MnO in olivine. The high Mn/Fe ratios (~100 × solar) in some olivines, cannot be attributed to any systematic correlation between planetary Mn/Fe ratio and heliocentric distance (Papike 1998), as Rietmeijer (2002) postulated for IDP olivines, because the range in the inner solar system is only a factor of ~2 and CAIs are unlikely to have formed in the outer part of the solar system.

Enstatite whiskers and platelets in chondritic IDPs provide structural and morphological evidence for an origin as condensates (Bradley et al. 1983). Enstatite whiskers contain axial screw dislocations characteristic of vapor phase growth (and some less plausible mechanisms). The abundance of the enstatite whiskers and platelets suggests they formed in the solar system.

Some magnesian matrix silicates may have condensed in the CAI-forming region close to the protosun as trace amounts of micrometer-sized grains that are refractory or $^{16}O$-rich are present in chondrite matrices (Choi et al. 1999; Kunihiro et al. 2004). However, most matrix silicates are $^{16}$O-poor, and several arguments discussed below suggest that crystalline matrix silicates were thermally processed in the region where chondrules formed.

3.4. *Relationship between Chondrite Matrices and Chondrules*

A number of chemical, isotopic and mineralogical features suggest that matrix silicates formed predominantly near chondrules. As noted above, the similarity of the oxygen isotopic composition of matrix and chondrules suggests that most matrix grains, including crystalline and amorphous silicates, formed in the $^{16}$O-poor region where chondrules formed and not where CAIs were produced. Several authors have also argued that matrix and chondrules are chemically complementary ensuring that the bulk chondrite composition is near solar (e.g., Wood 1996). In some cases, the argument is not very convincing . For example, Fe/Si ratios range widely among chondrites from 0.4-2 × CI chondrite (and solar) values, but for refractory elements, which are less fractionated, a stronger case can be made. Matrix and chondrules in a CR chondrite have mean Ti/Al ratios that are, respectively, 0.5 and 1.5 × the bulk ratio Ti/Al ratios (Klerner & Palme 2000). Since bulk chondrite values are all within 10% of the CI value, this implies that matrix and chondrules formed in the same nebular region where the refractory elements were vaporized and condensed.

The matrix in Kakangari, which differs from matrices in other chondrites, is closely related to its unique chondrules on the basis of bulk chemistry, mineralogy, and thermal history. The Kakangari chondrules clearly formed from material that closely resembled the associated matrix, and the matrix particles formed close to the chondrules. These features preclude formation of matrix silicates at large distances from the chondrules.

The thermal history of silicate crystals in chondrite matrices provides another link with chondrules. Some low-Ca pyroxene crystals in chondrite matrices cooled from above 1300 K at ~1000 K/h, a rate comparable to that experienced by chondrules, 10-1000 K/h (e.g., Zanda 2004). This suggests that the heating process that vaporized dust to form matrix grains also melted grain aggregates to form chondrules. Shocks, which are the most promising mechanism for melting mm-sized aggregates, also vaporize dust (Desch & Connolly 2002). To account for the presence in matrix material of many volatile elements at near-solar concentrations, it seems unlikely that matrix silicate dust condensed close to the protosun, as Shu et al. (1996) suggested. Matrix silicates probably formed under low ambient temperatures >2 AU from the protosun near chondrules.

The absence of melted matrix particles and the chemical uniformity of rims around different types of inclusions and chondrules show that matrix grains accreted together to form rims long after the chondrules cooled. Mineral compositions in igneous rims on many types of chondrules show that the material that accreted to chondrules and was later melted, or was partly molten when it accreted, was chemically similar to the enclosed chondrules (Krot & Wasson 1995). In the case of the low-FeO chondrules (type I), the igneous rims are made of low-FeO silicates and may largely have been derived from material that resembled the magnesian silicate component of matrix. For the FeO-rich chondrules (type II), the igneous rims are made of ferrous silicates and may have been derived largely from dust that resembled the amorphous component.

In Figure 2, we show schematically how the diverse silicate and refractory oxide components in chondrite matrices may have been formed and assembled. We infer that crystalline magnesian silicates and ferrous amorphous silicates condensed in regions where chondrules formed. Dust from the various chondrule-forming regions was mixed with small amounts of pristine materials that were not shock-heated significantly: presolar grains, organics, and Ca-Al-rich dust that formed in the nebular disk, presumably in the innermost regions with CAIs (Choi et al. 1999; Wood 2004). This mixture of materials accreted onto chondrules and refractory inclusions to form matrix rims. The presolar materials and organics may have escaped shock heating because they were located in the low-pressure periphery of the disk. Alternatively, they may have survived because they were derived from meter-sized icy bodies that drifted rapidly into the inner nebula where the ice evaporated (Cuzzi & Zahnle 2004).

3.5. *Further Implications*

Crystalline silicates, which appear to be nearly absent (≤0.2%) in the diffuse interstellar medium (Kemper et al. 2004), are abundant in long-period comets (up to 30-50%; Wooden et al. 2004; Honda et al. 2004a). This is generally attributed to thermal equilibration of amorphous silicate precursors in the hot inner regions of the solar nebula and transport outwards by turbulent diffusion (Hill et al. 2001; Bockelée-Morvan et al. 2002; Thompson et al. 2003; Gail 2004). Although substantial mixing between the hot inner region and the cool exterior of the solar nebula probably was necessary to mix traces of refractory dust from the CAI-formation region into chondrite matrices and IDPs (Fig. 2), we infer that many crystalline silicates in IDPs formed formed by condensation in localized heating processes, like matrix grains in chondrites. The solar nebular shocks that were invoked by Harker & Desch (2002) to form magnesian silicates at 5-10 AU by annealing amorphous silicates may also have vaporized and recondensed silicates. Forsterite crystals and amorphous silicate grains are present around T-Tauri and higher mass Herbig Ae/Be stars (Honda et al. 2003; Bouwman et al. 2003) suggesting that thermal processing of silicates was ubiquitous in protostellar disks.

There is no firm evidence from comets or chondrite matrices that crystalline, ferrous silicate dust formed in the solar nebula. There are rare clasts containing ferrous silicates in chondrites, which came from altered asteroids (see Scott & Krot 2003), but dust from such sources was evidently a very minor component during asteroid accretion. Crystalline ferrous silicates have not been observed around T-Tauri and Herbig Ae/Be stars. Crystalline ferrous silicates appear to be present around a Vega-like star, and these may be derived from colliding asteroidal bodies (Honda et al. 2004b). Ferrous silicates in chondritic IDPs, which are less abundant than magnesian silicates (Bradley 2003), may also come from asteroids, and, perhaps, from melt formed during atmospheric entry (Rietmeijer 2002). Chondrules rich in ferrous silicate (type II chondrules) probably formed from aggregates composed largely of amorphous particles of ferrous silicate that were melted.

We thank Jeff Cuzzi, Adrian Brearley, Joe Nuth and many other colleagues for valuable discussions and John Bradley for a helpful review. This work was partly supported by NASA grants NAG5-11591 (K. Keil) and NAG5-12882 (A. N. Krot).

Figure captions:

Fig. 1—Sketch based on transmission electron microscope images of Brearley (1993) and Greshake (1997) showing the mineralogy of matrices in the three pristine carbonaceous chondrites: fo: forsterite; en: enstatite; s: iron sulfide; i: iron-nickel. Rounded and elongated crystals of these minerals are embedded in an amorphous Fe-Mg silicate matrix as isolated crystals and clusters or aggregates of crystals.

Fig. 2—Schematic diagram showing how presolar silicate dust, which is largely amorphous ($\geq$99.8%), may have been thermally processed in the nebular disk before accretion into planetesimals. Dust that accreted at 2-3 AU into chondrite matrices contains traces of refractory dust from the inner edge of the disk ($\sim 10^{-5}$ to $10^{-4}$) and presolar crystalline silicates and refractory oxides ($\sim 10^{-5}$) and amorphous silicates ($\sim 10^{-6}$), which escaped thermal processing. These grains may have been dispersed by nebular turbulence or disk winds. Most chondritic matrix dust particles are crystalline, magnesian silicate and amorphous Fe-Mg silicate that condensed when dust aggregates were melted in the nebula to form chondrules, probably as a result of shocks. Dust that accreted into comets at >10 AU contains traces of refractory nebular dust ($\sim 10^{-6}$ to $10^{-5}$) and larger amounts of presolar, crystalline silicates and oxides ($\sim$0.1%). Most crystalline silicates in comets formed by condensation and annealing in the nebula; amorphous silicate particles may be presolar or nebular in origin. Contributions from planetesimal collisions may have dominated at later stages.

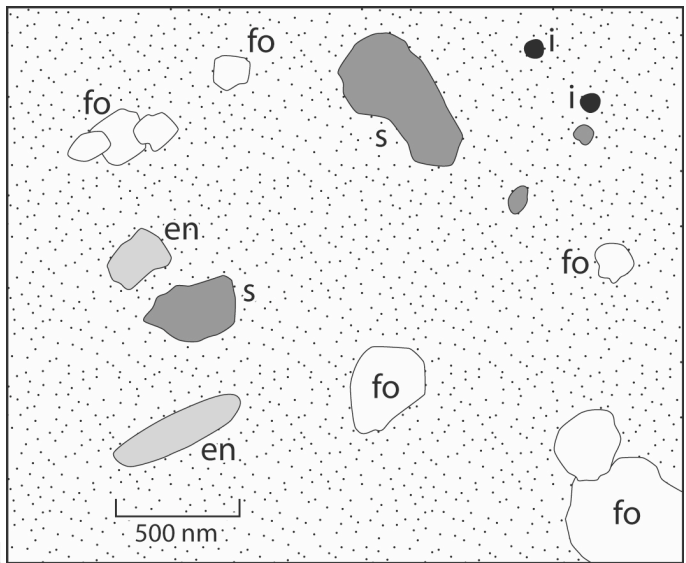

Fig.1

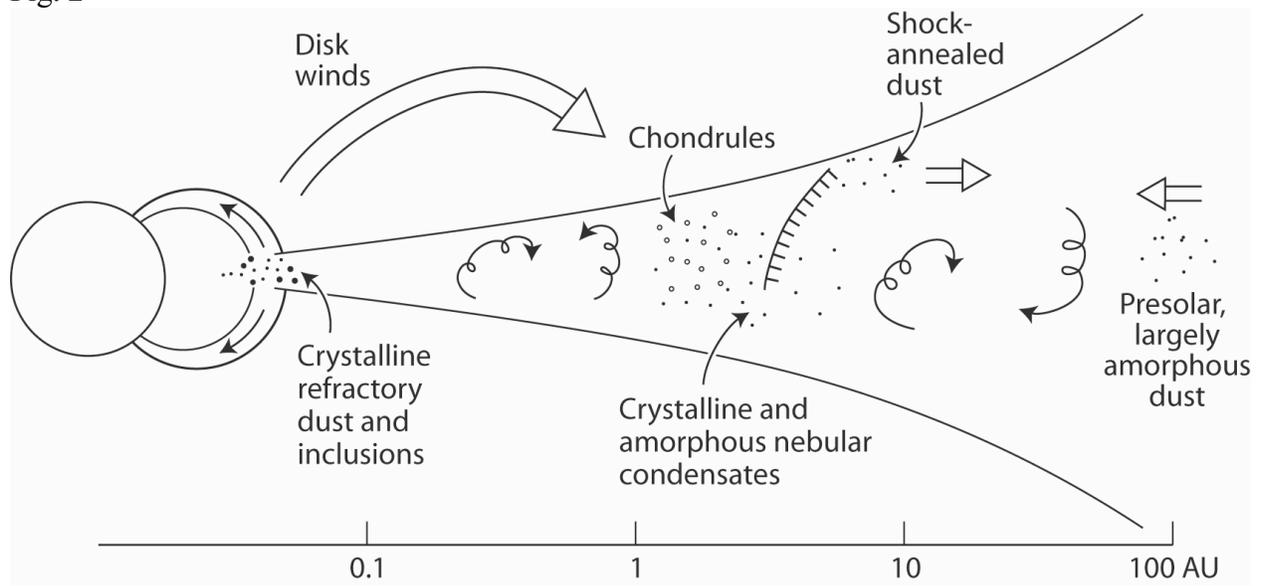

Fig. 2

Table 1. Minerals in chondrite matrices.

| Chondrite Group | Type | Minerals in matrix | Ref. |
|---|---|---|---|
| *Carbonaceous* | | | |
| CI | 1 | Serpentine, saponite, ferrihydrite, magnetite, Ca-Mg carbonate, pyrrhotite[1] | 1 |
| CM | 2 | Serpentine, tochilinite, pyrrhotite, amorphous silicate, calcite[1] | 1 |
| CR | 2 | Olivine, serpentine, saponite, minor magnetite, FeS, pentlandite, pyrrhotite, calcite | 2 |
| CO (ALHA77307) | 3.0 | Amorphous silicate, forsterite, enstatite, Fe,Ni metal, sulfides, Fe-silicates, magnetite | 3 |
| CO | 3.1-3.7 | Ferrous olivine, amorphous silicate, phyllosilicates, ferric oxide | 4 |
| CV reduced | 3 | Ferrous olivine, low-Ca pyroxene, low-Ni metal, FeS | 5, 6 |
| CV oxidized: | | | |
|   Bali-like | 3 | Ferrous olivine, phyllosilicate, fayalite, Ca-Fe pyroxene, pentlandite, magnetite | 7 |
|   Allende-like | 3 | Ferrous olivine, Ca-Fe pyroxene, nepheline, pentlandite | 8 |
| Ungrouped C: | | | |
|   Acfer 094 | 3.0 | Amorphous silicate, forsterite, enstatite, pyrrhotite, ferrihydrite, phyllosilicate | 9 |
|   Adelaide | 3 | Ferrous olivine, amorphous silicate, enstatite, pentlandite, magnetite | 10 |
| *Ordinary* | | | |
| LL (Semarkona) | 3.0 | Smectite, ferrous olivine, forsterite, enstatite, calcite, magnetite | 11 |
| H, L, LL | 3.1-3.6 | Ferrous olivine, amorphous silicate, pyroxene, albite, Fe,Ni metal | 4, 12 |
| *Other* | | | |
| K (Kakangari) | 3 | Enstatite, forsterite, albite, troilite, Fe,Ni metal, ferrihydrite | 13 |

References: ― (1) Zolensky et al. 1993; (2) Endress et al. 1994; (3) Brearley 1993; (4) Brearley & Jones 1998; (5) Lee et al. 1996; (6) Abreu & Brearley 2002; (7) Keller et al. 1994b; (8) Scott et al. 1988; (9) Greshake 1997; (10) Brearley 1991; (11) Alexander et al. 1989a; (12) Alexander et al. 1989b; (13) Brearley 1989.

Table 2. Properties of magnesian crystalline silicates and amorphous silicate in matrices of the least altered chondrites, chondritic IDPs, and Comet Hale-Bopp.

| Chondrite or Comet | Mg-rich Olivine | | | Mg-rich Pyroxene | | | Fe-rich Amorphous Silicate | |
|---|---|---|---|---|---|---|---|---|
| | Vol. (%) | Size (nm) | Fe/(Mg+Fe) | Vol. (%) | Size (nm) | Fe/(Mg+Fe) | Vol. (%) | Size (μm) |
| Acfer 094 | 30 | 200-300 | <0.02* | 20 | 200-400 | <0.03* | 40 | - |
| ALHA77307 | <20 | 200-4000 | 0.01-0.05* | <20 | 100-200 | <0.05?* | 50-80 | 1-5 |
| Adelaide | n.o. | | | ? | ~1000 | <0.02* | ~20? | - |
| Kakangari | 20 | 100-1000 | 0.01-0.03 | 50 | 100-1500 | 0.02-0.05 | n.a. | |
| Chondritic IDPs | - | 100-1000 | <0.05* | - | 100-1000 | <0.05* | - | 100-500 |
| Comet Hale-Bopp | 55 | ~200[+] | 0.1 | 15 | ~200[+] | 0.1 | 30 | ~200[+] |

n.o. not observed.
* Some grains are Mn-rich (0.6-2 wt.% MnO).
[+] Mean grain size of dust.
References: chondrites: Greshake (1997), Brearley (1993, 1991, 1989); chondritic IDPs: Bradley (2003), Reitmeijer (2002); Comet: Harker et al. (2002, 2004), Wooden et al. (1999).